\documentclass[12pt]{article}

\usepackage{amsmath}
\usepackage{graphicx,psfrag,epsf}
\usepackage{enumerate}
\usepackage{bbm}
\usepackage{caption}
\usepackage{subcaption}
\usepackage{makecell}
\usepackage{amssymb}
\usepackage{multirow}
\usepackage{xcolor}
\usepackage[scr=boondox]{mathalpha}

\usepackage[hidelinks]{hyperref}
\usepackage[linesnumbered,ruled,vlined]{algorithm2e} 
\usepackage{algpseudocode} 
\SetCommentSty{mycommfont} 
\SetKwInput{KwInput}{Input}                
\SetKwInput{KwOutput}{Output}              
\newcommand{\blind}{0} 
\date{} 
\usepackage[utf8]{inputenc}
\usepackage{bbm}
\usepackage{graphicx}
\usepackage{upgreek}
\usepackage[margin=.9in]{geometry}

\usepackage[backend=biber,style=numeric]{biblatex}
\usepackage{array}
\usepackage{hhline}
\newcolumntype{+}{!{\vrule width 2pt}}

\newlength\savedwidth

\newcolumntype{P}[1]{>{\centering\arraybackslash}p{#1}} 

\begin{document}

\def\spacingset#1{\renewcommand{\baselinestretch}%
{#1}\small\normalsize} \spacingset{1}

\if0\blind
{
  \title{Constructing Cell-type Taxonomy by Optimal Transport with Relaxed Marginal Constraints}

 \author{
  Sebastian Pena\textsuperscript{1}, Lin Lin\textsuperscript{2}, Jia Li\textsuperscript{1} \\
  \\
  \textsuperscript{1}Department of Statistics, The Pennsylvania State University \\ Email: \texttt{skp5799@psu.edu}, \texttt{jiali@psu.edu}\\
  \textsuperscript{2}Department of Biostatistics and Bioinformatics, Duke University \\ Email: \texttt{ll86@duke.edu}
}
  \maketitle
} \fi

\if1\blind
{
  \bigskip
  \bigskip
  \bigskip
  \begin{center}
    {\LARGE\bf Title}
\end{center}
  \medskip
} \fi

\maketitle

\bigskip
\section*{Abstract}
The rapid emergence of single-cell data has facilitated the study of many different biological conditions at the cellular level. Cluster analysis has been widely applied to identify cell types, capturing the essential patterns of the original data in a much more concise form. One challenge in the cluster analysis of cells is matching clusters extracted from datasets of different origins or conditions. Many existing algorithms cannot recognize new cell types present in only one of the two samples when establishing a correspondence between clusters obtained from two samples. Additionally, when there are more than two samples, it is advantageous to align clusters across all samples simultaneously rather than performing pairwise alignment. Our approach aims to construct a taxonomy for cell clusters across all samples to better annotate these clusters and effectively extract features for downstream analysis. A new system for constructing cell-type taxonomy has been developed by combining the technique of Optimal Transport with Relaxed Marginal Constraints (OT-RMC) and the simultaneous alignment of clusters across multiple samples. OT-RMC allows us to address challenges that arise when the proportions of clusters vary substantially between samples or when some clusters do not appear in all the samples. Experiments on more than twenty datasets demonstrate that the taxonomy constructed by this new system can yield highly accurate annotation of cell types. Additionally, sample-level features extracted based on the taxonomy result in accurate classification of samples.

\spacingset{1.5} 

\section{Introduction}
\label{secintro}
Many recent technologies, such as flow cytometry, Cytometry by time of flight (CyTOF), have led to the massive accumulation of single-cell data. Specifically, single-cell RNA sequencing data (scRNA-seq) has expanded from considering a single cell in 2009 to considering millions of cells by 2017~\cite{sve}. Advances in single cell technologies have allowed the measurement of global gene expression profiles of individual cells and have enabled the inspection of heterogeneous cell populations in complex samples that were inaccessible in bulk sequencing data. 

Numerous genes measured by scRNA-seq tools are uninformative and complicate downstream data analyses~\cite{Huang}. The general workflow  with scRNA-seq data involves reducing the number of genes, creating a manifold representation for visualization, identifying cell types by unsupervised clustering, and finding expression signatures of cell types by differential expression analysis~\cite{Peng}. While unsupervised clustering is crucial for discovering new cell types as well as unknown cell-gene relationships, manual annotation of these cell clusters is time-consuming and subjective. One major reason for the importance of cell categorization is that the proportion of each cell type in a sample and the associated differentially expressed marker genes can help predict the phenotype class of the sample. Example phenotypes include different levels of resistance to drugs treating cancer or different responses to vaccines, prognosis for diabetes and other diseases~\cite{Kim,Kotliarov,Segerstolpe,Xinjie,ruan2024pipet}. 

Next, we introduce some useful terminologies. The data we consider are those containing multiple, usually more than two, samples.  
Each sample is itself a dataset comprised of measurements on many cells. Specifically, a cell corresponds to a data point in the sample and the expression levels of genes are features measured for any cell. The same collection of genes, thus features, are considered for all cells in all samples.  However, certain cell types (i.e., particular groups of cells) may be present in one sample but not in the others. Usually, the samples are derived from multiple experiments on a single organism or distinct organisms within the same species. The cells in any sample are grouped into clusters, which correspond to different cell types, but the types may not have been identified. 

We aim to construct a {\it taxonomy} of clusters derived from all samples. Specifically, this taxonomy will be a hierarchical organization of clusters that identifies which clusters correspond to the same cell type and illustrates the relationships between different cell types in terms of their similarity. As clusters are often generated separately for individual samples, their labels are not coherent across samples. The central question we investigate here is how to establish a taxonomy to ensure consistent recognition of cell clusters. 

The taxonomy is valuable for multiple purposes. First, once a taxonomy is formed across samples, if one sample contains manually labeled cell types, the unlabeled cell clusters in other samples can be classified with known cell types. More detailed discussion on this usage scenario will be provided shortly. Secondly, the taxonomy can help us conduct downstream analysis. Consider the classification of samples according to their phenotype conditions, for instance, healthy patients versus those with a disease. With coherent cluster labels given to different samples, we can derive cluster-wise features such as cluster proportions and use them to classify samples. Note that we have regarded classification in two different contexts, that of clusters by their corresponding cell types and that of samples by their phenotype categories.

We emphasize the assumption that samples are not pooled for simultaneous processing. Our algorithm operates exclusively on the characteristics of clusters extracted independently from different samples, such as cluster proportions and cluster-wise average features, rather than the original data. This assumption precludes batch effect removal across samples (which often requires pooling) and synchronized clustering of data points across all samples.
The rationale for avoiding centralized analysis is multifaceted. Privacy concerns or the sheer size of the data may prohibit pooling. Additionally, clusters in different samples are often generated manually by expert researchers, and understanding the relationships between clusters across samples may be the primary objective. In such cases, pooling samples to cluster all data points together may not only be unnecessary but also meaningless.
While batch effect removal can contribute to constructing a taxonomy, and our algorithm demonstrates robustness to batch variations, the new algorithm is neither designed to remove batch effects nor operates under the same framework.

\section{Related Work} \label{ref:related}
The task of determining cell types from scRNA-seq data typically requires several cycles of grouping, sub-grouping, and merging of clusters. This procedure heavily depends on the manual input from subject matter experts 
who must balance the outcomes proposed by unsupervised clustering of multivariate data~\cite{kiselev2019challenges,kharchenko,Wang}. Despite efforts to streamline this process, cell types manually created continue to serve as the benchmark for most applications~\cite{Miao}. A common method for automatically assigning cell type labels to clusters involves utilizing reference data with expert verified ground truth labels~\cite{Alquicira,Hou,li2022neural, yang2022scbert, PredGCN, scMRA}. These reference clusters are matched with the unlabeled clusters. There are several statistical tools for matching cell clusters between different samples by quantifying the similarity of the gene expression profiles of cells in the clusters being compared~\cite{Peng}. This approach, however, has several drawbacks and negatively affects subsequent biological interpretations. It is noted that the presence of many cell types remarkably lessens the performance of classification tools~\cite{Peng}. Furthermore, the performance of classification is sensitive to the selection of the reference set~\cite{abde}. Finally, it is often a challenge to find a representative annotated training set encompassing all cell types present in the target datasets.

Cell labels for clusters are usually not available in published scRNA-seq data. One study found that nearly half of $72$ datasets did not include inferred cell types~\cite{Squair}. However, even when the biological nature of the cell clusters are unknown, coherent labels assigned to clusters across samples are highly useful. Consistent labels can help merge clusters across samples; and the integrated clusters can better indicate marker genes. 
These marker genes are usually chosen based on a specific computational clustering method and are utilized to annotate clusters~\cite{pullin}. 

Even in the presence of an annotated reference sample, automatic annotation methods have had limited success. Clusters that represent identical or similar cell types may have annotations suggesting otherwise.
Mislabeling can be caused by
the dropout phenomenon that ``a gene is observed at moderate or even high expression level in one cell but is not detected in another cell''~\cite{kharchenko} as well as batch effects~\cite{haghverdi, tung}. 
To the best of our knowledge, an automatic tool has not been developed to build a taxonomy of clusters in more than two samples. Instead, cluster labels in a reference sample are used to annotate clusters in other samples in a pairwise fashion~\cite{Hou,kis,yun,azad,Alquicira}. 
It is known that compared to bulk-cell data, scRNA-seq data has a higher level of noise due to both biological and technical reasons~\cite{Mou}. As a result, often times, no single sample serves as a good reference for annotating cell clusters. We expect that the approach of building a taxonomy of clusters based on all samples, in contrast to a pair of samples, can potentially mitigate the effect of noise and thus yield more accurate cluster annotation, an advantage confirmed by our experiments. 

By treating all samples simultaneously, our proposed method creates a hierarchy of clusters without using an annotated reference sample. Our method combines state-of-the-art optimal transport (OT) techniques and hierarchical clustering. 
\textit{ClusterMap}~\cite{gao} is an existing method with a similar logical framework as ours, which also considers multiple datasets with predefined clusters. Using marker genes, this method creates a hierarchy of clusters based on cluster similarity measured by the Jaccard distance. The main drawback to this approach is the necessity of marker genes for each dataset prior to implementing the algorithm. Marker genes are useful for annotating cells that belong to some major cell types. Different marker gene databases often list different sets of marker genes for the same cell population~\cite{abde}. Furthermore, the same marker genes frequently manifest in various cell clusters, and may represent diverse cell types~\cite{clarke, ianevski}. Lastly automatic methods that determine marker genes for clusters often disagree and are affected by the noisiness of scRNA-seq data~\cite{pullin,qiu}. 

OT has been increasingly used in machine learning and single-cell data analysis~\cite{Luis,lin2023multisource,zhang2023multi,zhang2022bsde,cao2022unified, Hui}. Although OT is a prominent framework for matching, there are other paradigms such as step-wise merging (similar to the way a dendrogram is formed) used by QFMatch~\cite{orlova} and bipartite graph partitioning (BGP) used by flowMatch~\cite{azad}. 
The formulation of the standard OT contains two marginal constraints which reflect the implicit assumption that the proportions of matched clusters are fixed (modulo randomness in observation) across all the datasets. This assumption is improper for scRNA-seq data where one or more cell types can be present in some samples but not others.~\cite{Jia} proposed a new formulation, namely, Optimal Transport with Relaxed Marginal Constraints (OT-RMC), by introducing gap variables to relax the marginal constraints. They have demonstrated that OT is too rigid for various matching problems. We adopt as a core technique OT-RMC and develop a new algorithm that builds a taxonomic tree made up of clusters from all samples. For comparison, we will also experiment with Partial Optimal Transport (POT) which relaxes standard OT by transporting only a fraction of the total mass \cite{benamou2015iterative,flamary2021pot}.

\section{The Approach}
\label{method}
\subsection{Framework for Constructing Taxonomy}\label{secframework}

\begin{figure*}[ht]
    \centering
    \includegraphics[width=0.9\textwidth]{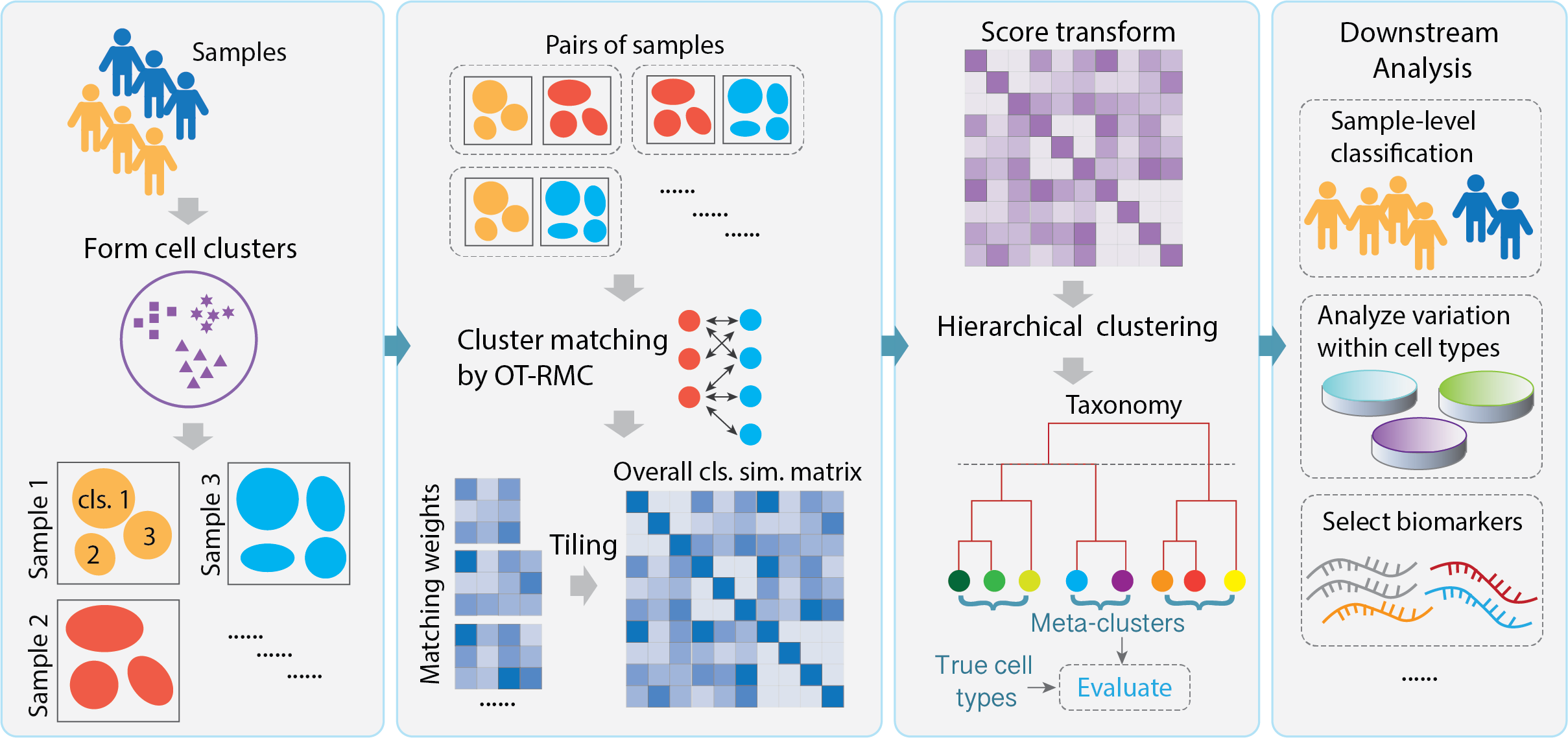}
{\spacingset{1}    \caption{A schematic representation of the Multisample OT Taxonomy (MOTT) system. The taxonomy hierarchically organizes clusters across samples, identifying which clusters correspond to the same cell type and illustrating the relationships between different cell types based on their similarity.}
    \label{Fig1}
    }
\end{figure*}

Consider a collection of samples $\mathcal{X}_i$, $i=1, ..., N$, each containing measurements on a set of cells. Specifically, $\mathcal{X}_i=\{x^{(i)}_1, ..., x^{(i)}_{m_i}\}$, where $x^{(i)}_{j}\in \mathbb{R}^d$ and $d$ is the dimension of the cell-wise feature vector. Suppose that each sample $\mathcal{X}_i$ has been clustered. Denote clusters generated for $\mathcal{X}_i$ by $C^{(i)}_k$, $k=1, ...,n_i$, where $C^{(i)}_k$ contains the points (aka, cells) assigned to the $k$th cluster. We refer to $C^{(i)}_k$'s as {\it cell clusters} or simply clusters if it is clear from the context. Note that $\mathcal{X}_i=\cup_{k=1}^{n_i}C^{(i)}_k$. Denote $\mathcal{P}_i=\{C^{(i)}_1, C^{(i)}_2, ..., C^{(i)}_{n_i}\}$. 

We address the problem of constructing a taxonomy for all clusters in all samples when the clusters $C^{i}_k$ from different samples are not labeled consistently. Inconsistent labeling can occur in various cases, the simplest being label permutation. However, in practice, a cluster in one sample may split into multiple clusters in another sample, or conversely, multiple clusters may merge into one. Additionally, a new cluster may emerge. Moreover, due to sample variations and nuances in clustering, determining relationships between clusters across samples is often challenging.
We propose a system to build a taxonomy tree that reveals which clusters in different samples represent the same cell-type and how the cell-types relate to each other in a hierarchy of similarity. We employ techniques of OT and dendrogram clustering to effectively leverage information across all clusters in one sample and across all samples. We refer to the system as {\it Multisample OT Taxonomy} ({\it MOTT}).  A schematic flowchart for MOTT is provided in Fig~\ref{Fig1}.

A taxonomy constructed by MOTT can help us understand relationships between cell types that are not manually annotated and allow for coherent labeling of cell clusters across samples, which is valuable for downstream analysis.
Next, we describe the main components in the MOTT system, as illustrated in Fig~\ref{Fig1}. 

\begin{enumerate}
    \item For every pair of samples $\mathcal{X}_i$ and $\mathcal{X}_j$, $i, j\in \{1, ..., N\}$, compute the cost of matching cluster $C^{(i)}_k$ and $C^{(j)}_l$, denoted by $D(C^{(i)}_k, C^{(j)}_l)$, $k=1, ..., n_i$, $l=1, ..., n_j$. We refer to the $n_i\times n_j$ matrix containing entries $D(C^{(i)}_k, C^{(j)}_l)$ as the cost matrix, which is denoted by $\mathbf{D}_{i,j}$. Apply OT-RMC to compute a matching weight matrix of size $n_i\times n_j$, denoted by $\mathbf{W}_{i,j}=(w^{(i,j)}_{k,l})_{k=1, ..., n_i, l=1, ..., n_j}$ based on the input cost matrix $\mathbf{D}_{i,j}$. We only need to compute $\mathbf{W}_{i,j}$, $j\geq i$, since $\mathbf{W}_{j,i}=\mathbf{W}_{i,j}^T$. The weights $w^{(i,j)}_{k,l}\in[0,1]$ indicates the extent of matching cluster $C^{(i)}_k$ in sample $\mathcal{X}_i$ with cluster $C^{(j)}_l$ in sample $\mathcal{X}_j$, a higher value corresponding to stronger matching. We then compute a normalized version of $\mathbf{W}_{i,j}$, denoted by $\widetilde{\mathbf{W}}_{i,j}$, by performing row-wise and column-wise normalization on $\mathbf{W}_{i,j}$ and then computing the average of the two matrices. We treat the matching weights in $\widetilde{\mathbf{W}}_{i,j}$ as the similarity scores between clusters. The detailed formulation of OT-RMC is explained in the next subsection. 
    
    \item Form matrix $\mathbf{B}$ by tiling $\widetilde{\mathbf{W}}_{i,j}$, $i=1, ..., N$, $j=1, ..., N$. Let $n=\sum_{i=1}^{N} n_i$. $\mathbf{B}$ is an $n\times n$ matrix that contains $\widetilde{\mathbf{W}}_{i,j}$ as its $(i,j)$th block.
  We refer to $\mathbf{B}$ as the {\it overall cluster similarity matrix}, which contains the similarity score between any pair of clusters across all samples. To convert the similarity matrix $\mathbf{B}$ into an {\it overall cluster distance matrix} $\mathbf{A}$, we apply an entry-wise transform $-\log(\cdot)$ and then standardize the matrix by dividing each entry by the maximum entry. The nonlinear transform $-\log(\cdot)$ amplifies the difference between matching weights near $0$, which occur frequently, thereby allowing us to better capture cluster distances. For any entry smaller than $\xi\cdot 10^{-7}$, where $\xi$ is the smallest cluster proportion in the dataset, instead of applying $-\log(\cdot)$ transform, we set the entry to $7-\log(\xi)$. This modification ensures that we never encounter $\log(0)$ and mitigates the impact of precision errors in the solution of OT. Experimentally, it was found that the similarity scores in $\mathbf{A}$ smaller than $10^{-10}$ have the same impact on building the taxonomy.

    \item 
  We apply dendrogram clustering using Ward's linkage method with the
\textit{linkage()} 
function in \textit{Matlab}, using $\mathbf{A}$ as the input pairwise distance matrix. The result is the taxonomy tree (a dendrogram). We then apply the \textit{cluster()} function
to the taxonomy tree to generate the final grouping of clusters. The number of groups is set equal to the number of ground truth cell types in the original dataset. 
\end{enumerate}

To summarize, our final result is a partitioning of the cell clusters across all samples. Clusters assigned to the same group are considered to represent the same cell type. We call a group of clusters obtained by a taxonomy a {\it meta-cluster}. The MOTT system aligns clusters across all samples in a holistic manner, an alignment approach which we refer to as  {\it Simultaneous Alignment} (SA) in the discussion below.  
We can assign distinct labels to each group of clusters, thus generating consistent cluster labels across samples. Although our method does not explicitly address this, we have never observed instances where different clusters from the same sample are grouped together, as the large distances between different clusters within the same patient generally prevent this. 


\subsection{Matching Clusters by OT-RMC} \label{secot}
Consider two samples with partitions $\mathcal{P}_i=\{C^{(i)}_1, ..., C^{(i)}_{n_i}\}$, $i=1, 2$, where $C^{(i)}_k$ is the $k$th cluster in the $i$th partition and $n_i$ is the number of clusters. 
The distance matrix $\mathbf{D}$ contains entries $D(C^{(1)}_k, C^{(2)}_l)$, representing the distance between any pair of clusters across the two partitions.
To determine the matching weights between clusters in the two samples, we employ OT-RMC, formulated as follows.
Let {\it gap vector} $\mathbf{g}_1=(g_{1,1}, ..., g_{1, n_1})^T$ and $\mathbf{g}_2=(g_{2,1}, ..., g_{2, n_2})^T$. Let $\mathbf{g}$ be the concatenation of $\mathbf{g}_1$ and $\mathbf{g}_2$. 
Roughly speaking, the gap vector $\mathbf{g}$ 
indicates the extent of deviation from the marginal constraints, $\mathbf{q}_1$ and $\mathbf{q}_2$, in OT (a review of basic OT is provided in Section 1 in \textbf{S1 Appendix}). Nonzero values in $\mathbf{g}$ are penalized to ensure that OT-RMC does not degenerate into trivial solutions. Depending on the penalty function used for $\mathbf{g}$ and whether and how upper and lower bounds are set up for the variables in the optimization problem, OT-RMC can have many variations. In our case here, the penalty for the gap variables is the $L_1$ norm denoted by $L(\mathbf{g})$. The optimization problem is presented in Problem (\ref{eq:OT-RMC1}) below.
\begin{eqnarray}
R(\mathbf{D},\mathbf{q}_1,\mathbf{q}_2)&= \min_{\mathbf{W},\mathbf{g}} \langle\mathbf{D},\mathbf{W} \rangle +\lambda L(\mathbf{g})\\
\textrm{s.t.} \quad 
  &\mathbbm{1}^T_{n_{1}}\cdot \mathbf{W}\cdot\mathbbm{1}_{n_{2}}=1 \, , \quad
  \mathbf{W}\geq0 \\
  &\mathbf{q}_1-\mathbf{g}_1\leq \mathbf{W} \cdot \mathbbm{1}_{n_{2}} \leq \mathbf{q}_1+\mathbf{g}_1\\
  &\mathbf{q}_2-\mathbf{g}_2\leq \mathbf{W}^T \cdot \mathbbm{1}_{n_{1}} \leq \mathbf{q}_2+\mathbf{g}_2\\
  \label{eq:OT-RMC1}
\end{eqnarray}

Denote the optimal $\mathbf{W}$ for Problem (\ref{eq:OT-RMC1}) above by the matching weight matrix $\mathbf{W}^{*}$. Matrix $\mathbf{W}^{*}=(w^{*}_{k,l})$ specifies the matching proportion assigned to each pair of clusters $C_k^{(1)}$ and $C_{l}^{(2)}$ after optimizing the matching cost. The proportion $\tilde{q}^{(1)}_k=\sum_{l=1}^{n_2}w^{*}_{k,l}$ is called induced proportion given to $C^{(1)}_k$ and $\tilde{q}^{(2)}_l=\sum_{k=1}^{n_1}w^{*}_{k,l}$ is the induced proportion given to $C^{(2)}_l$. If it is overly costly to match a cluster with any cluster in the other partition, this cluster can be assigned with zero proportion, that is, $\tilde{q}^{(1)}_k=0$ or $\tilde{q}^{(2)}_l=0$. 

To define the distance $D(C^{(1)}_k, C^{(2)}_l)$, we first compute the squared Wasserstein distance between two Gaussian distributions characterizing clusters $C_k^{(1)}$ and $C_l^{(2)}$ respectively. Suppose the Gaussian distribution representing $C_k^{(i)}$ is $N(\mu^{(i)}_k, \Sigma^{(i)}_k)$, where $\mu^{(i)}_k$ is the mean vector and $\Sigma^{(i)}_k$ is the covariance matrix. Then
\begin{align*}
    D^2_W(C^{(1)}_k, C^{(2)}_l) &= ||\mu^{(1)}_k-\mu^{(2)}_l||^2_2 + 
    \mathbf{Tr}\left[\Sigma^{(1)}_k+\Sigma^{(2)}_l-2\left( \Sigma^{(1) \frac{1}{2}}_k \Sigma^{(2)}_l\Sigma^{(1) \frac{1}{2}}_k  \right)^{\frac{1}{2}}\right] \,.
\end{align*}
For any cluster, we estimate the mean $\mu^{(i)}_k$ by its sample mean 
and $\Sigma^{(i)}_k$ by its sample covariance matrix. We then standardize $D_W(C^{(1)}_k, C^{(2)}_l)$ by dividing each entry indexed by $k=1, ..., n_1$ and $l=1, ..., n_2$ by the maximum entry. Define the standardized matrix as the cost matrix $\mathbf{D}=(D(C^{(1)}_k, C^{(2)}_l))_{k=1, ..., n_1, l=1, ..., n_2}$. 
The standardization used to define $\mathbf{D}$ makes $\lambda$ less dependent on the data. How to select $\lambda$ in our experiments is explained in the Results section.


\subsection{Taxonomy based on Alignment with a Reference} \label{secreference}

In this subsection, we introduce an alternative approach to constructing a taxonomy based on alignment with a pre-chosen reference. We refer to this method as {\it Reference Alignment} (RA). 

The standard approach for labeling clusters in a sample involves pairwise matching~\cite{Hou,kis}. With pairwise matching, clusters in one sample are treated as the ground truth, and clusters in another sample are labeled by aligning them with the clusters in the reference sample. The sample providing the cluster labels is called the reference. By aligning the clusters of every other sample with those of the reference sample, coherent labels can be assigned across all samples, thereby constructing a taxonomy. This process can be repeated with different reference samples. Specifically, we implement the following process to provide a baseline for comparison with the MOTT system.

The samples chosen as references are those with the highest number of unique cell types. A reference sample may not contain all the ground truth clusters present in the other samples. We then use a reference sample to label the clusters of the rest of the samples. Suppose the reference sample has $n_1$ clusters and the sample to be aligned has $n_2$ clusters. OT, OT-RMC, and POT all yield a $n_1 \times n_2$ weight matrix, $\mathbf{W}$. According to the way we select a reference sample, $n_1 \geq n_2$. The assigned label for the $l$th cluster in the non-reference sample is determined by $k^{*} = \arg\max_k W_{k, l}$. 

For evaluation, we can repeat the process using different samples as the reference and measure the performance by averaging results obtained using different references.     

\subsection{Sample-level Classification}
\label{secsampleclassify}

We evaluate a taxonomy by comparing it with the true annotations of cell types, if provided in the data, and by computing the accuracy of classifying samples based on feature vectors extracted using the taxonomy. The latter is an indirect approach, assessing the effect of the taxonomy on a downstream analysis task. Once a taxonomy is established, we assign coherent labels to clusters in all samples. We then compute the proportions of cells belonging to each cluster label and use these proportions as features for classifying samples. Thus, the dimension of the feature vector corresponds to the number of distinct cluster labels. Note that even if the true types of cell clusters are unavailable, for instance, in the case of computationally generated clusters, an automatically generated taxonomy will still allow for coherent labeling of cell clusters and, consequently, enable sample-level classification.
In our experiments, we test different approaches to constructing the taxonomy, some serving as baselines for comparison, and apply Random Forest (RF) to classify samples.

\section{Results}
\label{secresults}

\subsection{Experimental Setups} \label{secexp}
We evaluate MOTT in terms of cell type accuracy for cell clusters and classification accuracy for samples. We tested it on 11 scRNA-seq datasets. Most datasets do not divide samples into categories or provide labels for all samples. Therefore evaluation by sample-level classification is conducted only on 4 datasets. For the tested datasets, different samples typically correspond to different patients, though in some cases, they correspond to replicates recorded at different times. The task of constructing a taxonomy becomes more interesting and challenging with a large number of samples. Unfortunately, many scRNA-seq datasets lack metadata to allocate cells to different samples. In other cases, while the dataset might include sample division information, the number of samples might be too few (2 to 5). To effectively examine our new method, we generated simulated partitions of cell-level data to create ``{\it simulated samples}'' (SS). Since simulated samples are not subject to batch effects (which commonly occur in scRNA-seq data), differences in results by various methods can be better attributed to the characteristics of the methods.


For baseline methods, we consider alternatives to MOTT from two perspectives: using OT and POT instead of OT-RMC for computing matching weights between clusters, and using RA instead of SA for aligning all clusters. Hence, we compared a total of six methods. If cluster matching by OT-RMC in MOTT is replaced by OT or POT, we name the method OT-SA and POT-SA respectively. If SA is replaced by RA, we name the method OT-RMC-RA, POT-RA, or OT-RA, depending on which cluster matching method is used. Finally, the method OT-RMC-SA corresponds to MOTT. The only natural baseline method from existing literature is OT-RA. However, since the novelty of MOTT comes from employing both OT-RMC and SA, we have investigated all six methods.

To apply MOTT, we start with the cell clusters provided for each sample. The 11 datasets we tested include ground truth cell types, which are used to form cell clusters within each sample. To perform cluster matching using OT or OT-RMC, as explained in the previous section, we estimate a Gaussian distribution in a reduced-dimensional space for each cluster based on all the cells it contains. To reduce dimension, we first select highly variable genes using the R tool \textit{HVGs}. We then apply Principal Component Analysis (PCA) and select the number of principal components needed to account for $90\%$ of the variability.

\subsection{Evaluation Measures} \label{secevaluate}
Suppose a taxonomy for cell clusters has been established, specifically generating meta-clusters. All clusters in the same meta-cluster are regarded as coming from the same cell type. To compare with the ground truth, we assign a cell type to each meta-cluster by majority vote. In particular, the most common cell type among the clusters in a meta-cluster is taken as the type of the meta-cluster. Ideally, if all clusters in one meta-cluster are of the same type, the taxonomy would have consistently labeled them. Otherwise, errors occur in labeling the cell types of some clusters. 

We then compute the accuracy of labeling cell clusters in two ways. The cluster-level accuracy, denoted by $\zeta_{cls}$, is defined as the percentage of clusters (across all samples) for which the ground truth cell type matches the cell type indicated by the meta-cluster. The cell-level accuracy, denoted by $\zeta_{cell}$, is defined as the percentage of cells that are assigned with correct cell types based on the meta-clusters. Essentially, $\zeta_{cell}$ is a weighted average version of $\zeta_{cls}$, where the weight of a cluster is proportional to the number of cells contained in it, meaning that larger clusters have a greater impact on the overall accuracy calculation. 

In addition, we compute the Adjusted Rand Index (ARI) for cell clusters based on ground truth cell labels and the partition of cells based on meta-clusters. While calculating the ARI, cells from all samples are pooled together. Specifically, cells from all samples with the same ground truth cell labels or meta-cluster labels are grouped into the same cluster.

\subsection{scRNA-seq Datasets}

\begin{figure*}[!htb]
\centering
\begin{tabular}{c}
\begin{tabular}{cc}
  \includegraphics[width=2.4in]{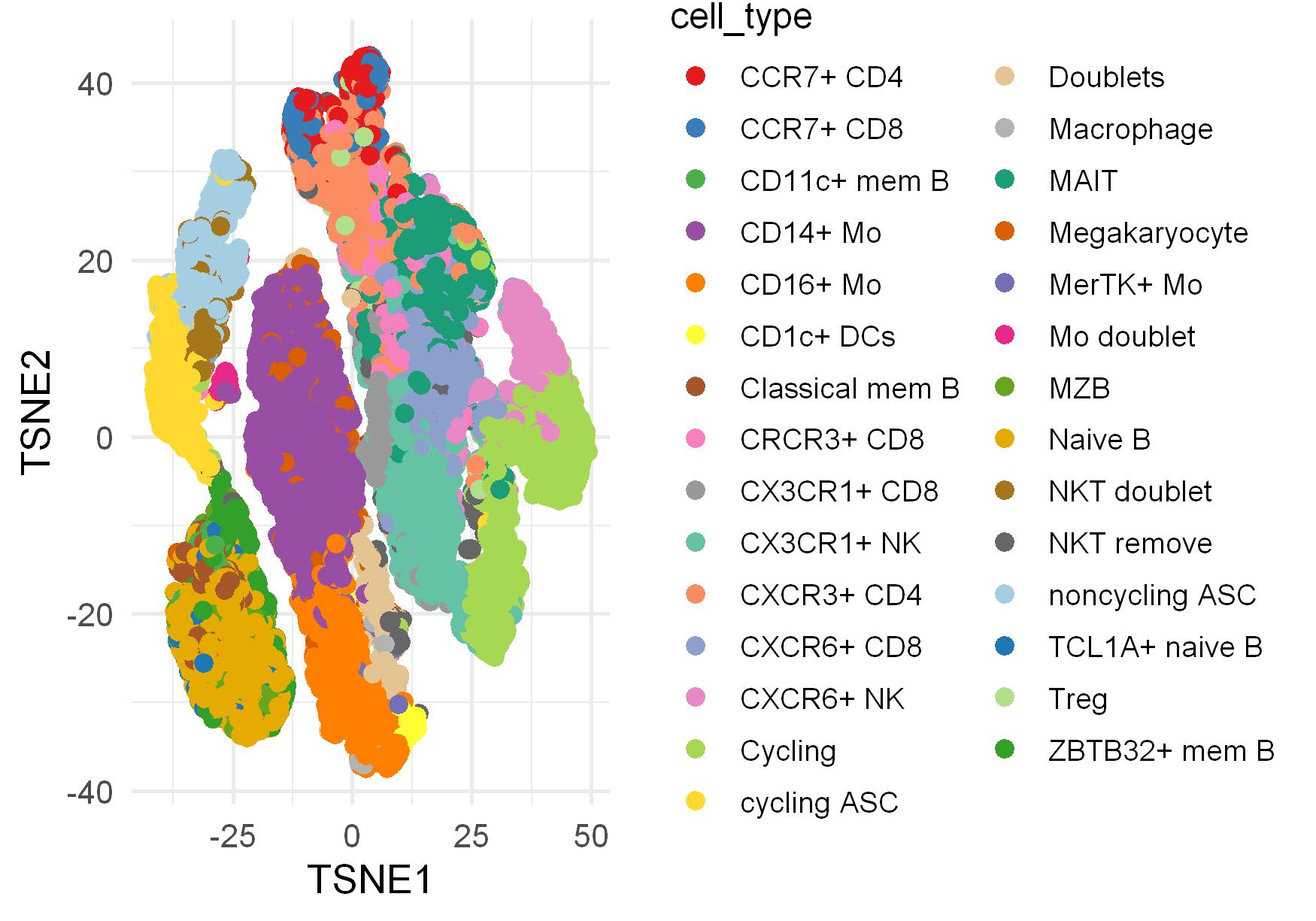}
&
  \includegraphics[width=2.35in]{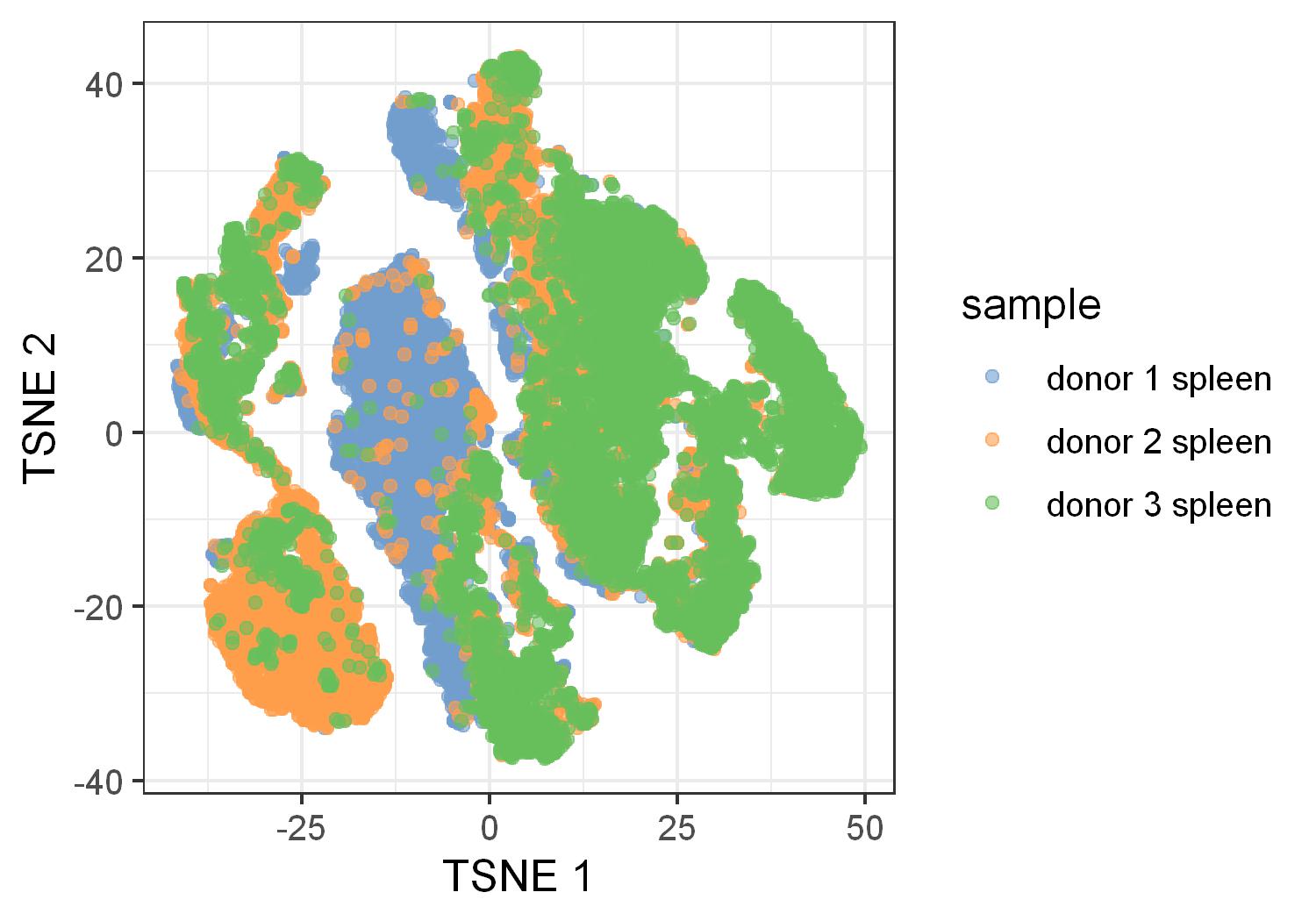}
  \end{tabular} \\
(a)\\ \\
\begin{tabular}{cc}
  \includegraphics[width=2.4in]{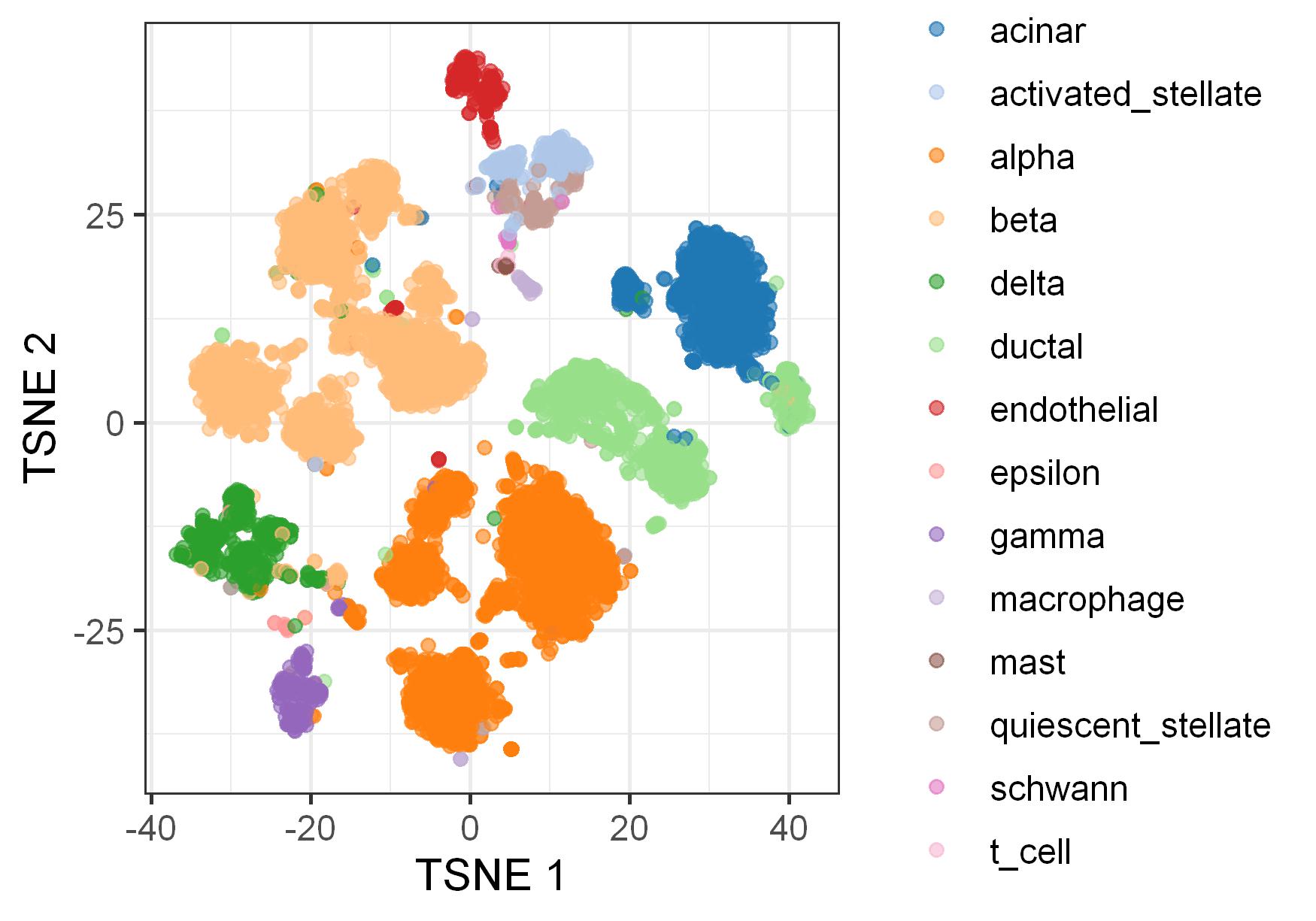}
&
  \includegraphics[width=2.35in]{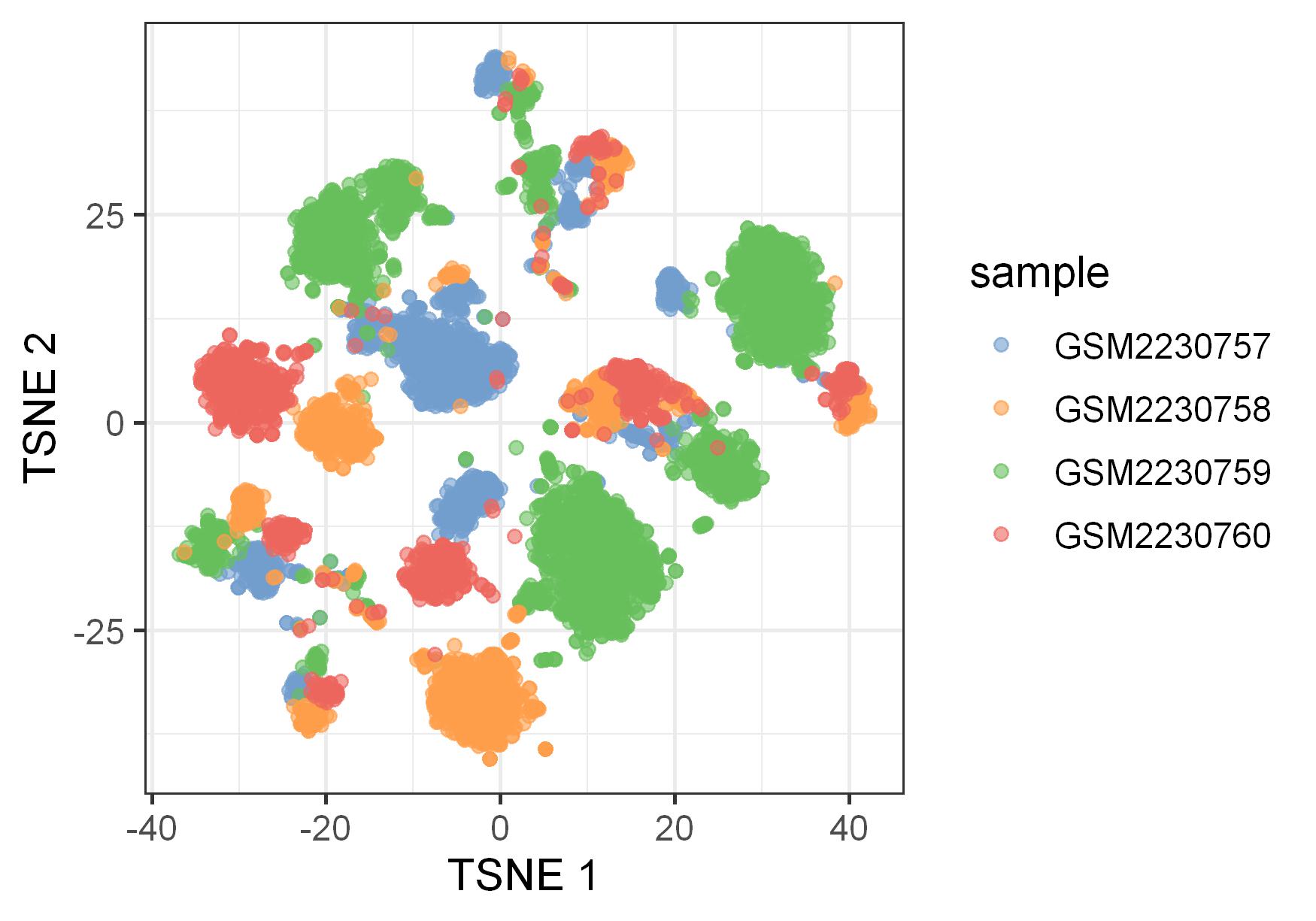}
  \end{tabular} \\
(b)\\ \\
\end{tabular}
{\spacingset{1}
\caption{t-SNE plots showing data grouped by ground truth cell types (left column) and by sample (right column). (a) Zhao Spleen dataset, (b) Baron Pancreas dataset. The Zhao Spleen dataset, which only includes cells from the spleen, was used instead of the entire Zhao dataset because the latter contains over $60,000$ points, making it too large for effective visualization. The composition of cell types in different samples, both in terms of existing cell types and their population sizes, varies substantially. This observation highlights the challenge of constructing a taxonomy for cell clusters across samples.}
\label{Fig2}
}
\end{figure*}

We used 11 data sources: {\it Segerstolpe}~\cite{Segerstolpe}, {\it Bacher}~\cite{bacher}, {\it Baron}~\cite{baron}, {\it Zhao}~\cite{zhao}, {\it Tasic}~\cite{tasic}, {\it Zilionis Human}~\cite{zilionis}, {\it Zilionis Mouse}~\cite{zilionis}, {\it Lawlor}~\cite{lawlor}, {\it He Organ}~\cite{He}, {\it Jessa Brain}~\cite{jessa}, and {\it Wu}~\cite{Wu}. Two datasets were formed from the Jessa Brain data based on different regions of the brain (pons versus forebrain). 
Table~\ref{datasets} provides basic information about the datasets.
Notably, two datasets, Zhao and Zillions Human, include two sets of ground truth labels: one coarse (representing major cell types) and one fine (representing cell subtypes). We experimented with both sets of ground truth labels. More detailed descriptions of the datasets and data preprocessing are provided in Section 2 of \textbf{S1 Appendix}. Fig~\ref{Fig2} presents a pair of t-SNE plots for two example datasets, where the data points are colored according to their ground truth cell types and the samples they belong to. The plots show the complexity of the datasets, as the cell population sizes vary significantly and batch effects are present in both.

 As aforementioned, some datasets do not divide cells into samples. For this reason, and to facilitate performance comparisons, we generated simulated samples by randomly splitting the entire dataset. Unless specified otherwise, an original dataset was randomly partitioned into 20 subsets, each treated as one sample. For the Segerstolpe, Zilionis Mouse, and Wu datasets, cells were collected under multiple conditions, and we generated samples separately by randomly dividing cells under each individual condition. We refer to such simulated samples as ``simulated by label'' (SL). Note that the cell condition is different from the cell type; instead, the cell condition corresponds to the class label assigned to each sample.  

Sample-level classification was conducted on the Segerstolpe, Bacher, Zilionis Mouse, and Wu datasets. Specifically, samples in Segerstolpe were labeled in the original data with two categories: patients with type II diabetes mellitus and healthy donors (4 versus 6). Similarly, the Bacher samples were originally labeled as patients previously exposed to a virus or those unexposed (14 versus 6). For the Zilionis Mouse and Wu datasets, sample division was not indicated in the original data, but cells come from two conditions: healthy versus lung tumor or diseased. Simulated samples were generated by randomly dividing either diseased or healthy cells (the SL scheme), and the samples were thus labeled accordingly as either diseased or healthy (10 versus 10). Additionally, the Wu dataset lacks ground truth cell types; instead, hierarchical clustering into 10 clusters was used to simulate cell types. 

\begin{table}[htbp]
\centering
\begin{tabular}{|p{2.4cm}|P{0.7cm}|P{1.4cm}|P{1.4cm}|P{1.4cm}|P{0.7cm}|}
 \hline
 Dataset & Smp & C. Types & Cells & Genes & PCs\\
 \hline
 Segerstolpe &   10  & 11 & 2,056 & 24,454 & 50\\
 Bacher &20 & 6  & 104,417 & 33,538 & 50\\
 Baron &4 & 14 & 8,569 & 18,845 & 50\\
 Zhao    &9 & 4 \& 29  & 60,672 & 33,694 & 8 \\
 Tasic&   -  & 7&  1,727& 24,058  & 50 \\
 Zilionis Human& 7  & 21 \& 35 & 54,773  & 41,861 & 50 \\
 Zilionis Mouse& -  & 6 & 15,905  & 28,205& 50 \\
 Lawlor&  8 & 7 & 617 & 26,616& 50 \\
 He Organ & -  & 42 & 76,037 & 12,021  & 20\\
 Jessa Pons& -  & 27& 27,954 & 27,998 & 2 \\
 Jessa Forebrain & -  & 27& 33,641 & 27,998 & 2 \\
 Wu & - & - & 18,249 & 17,542 & 9 \\
 \hline
\end{tabular}
{\spacingset{1} \caption{Summary information on the scRNA-seq datasets. 
For each dataset, we list the numbers of samples, cell types, cells, genes, and principal components (PCs) retained in the analysis. For some datasets, although cells might have been collected from different donors or organs, information on how to divide the cells into samples is lacking. In such cases, we do not list the number of samples. For two of the datasets, cells are categorized at both a coarse and a fine level, with more cell types annotated at the fine level.}
\label{datasets}
}
\end{table}

\subsection{Accuracy of Taxonomy}
\label{secaccuracy}
\begin{figure*}[ht]
    \centering
    \includegraphics[scale =.09]{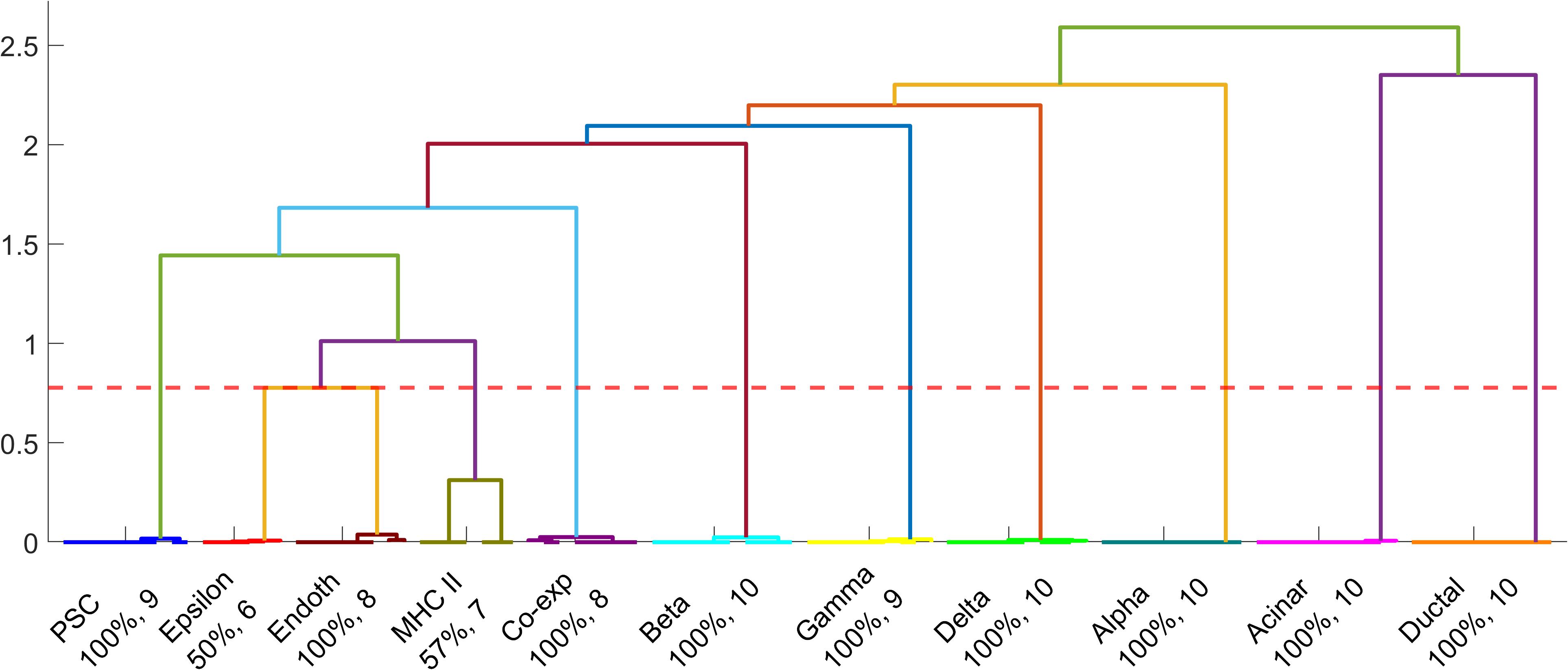}
    {\spacingset{1} \caption{An example taxonomy created by MOTT for the Segerstolpe dataset. The Segerstolpe dataset includes ten samples and, collectively, 11 cell types. Correspondingly we use the taxonomy to generate 11 meta-clusters. Each meta-cluster is labeled according to the most common cell type among its constituent clusters. Beneath each meta-cluster, we label its chosen cell type, and in the second line, show the number of clusters in this meta-cluster and the percentage of clusters assigned with the correct cell type. Cell type ``Endoth'' stands for Endothelial, ``Co-exp.'' stands for Co-expression, and ``MHC II'' stands for ``MHC Class II''. Only two meta-clusters, Epsilon and MHC Class II, include cluster cell types other than their chosen labels. The horizontal dashed line indicates the cut-off level at which eleven meta-clusters are formed. }
    \label{Fig3}
    }
\end{figure*}

As an example, Fig~\ref{Fig3} presents the taxonomy created by MOTT for all the ten samples from the Segerstolpe dataset. For clarity, we do not display results for all samples, as the resulting dendrogram would be too large.

\begin{figure}[p]
    \centering
\includegraphics[width=\textwidth,height=0.22\textheight,keepaspectratio]{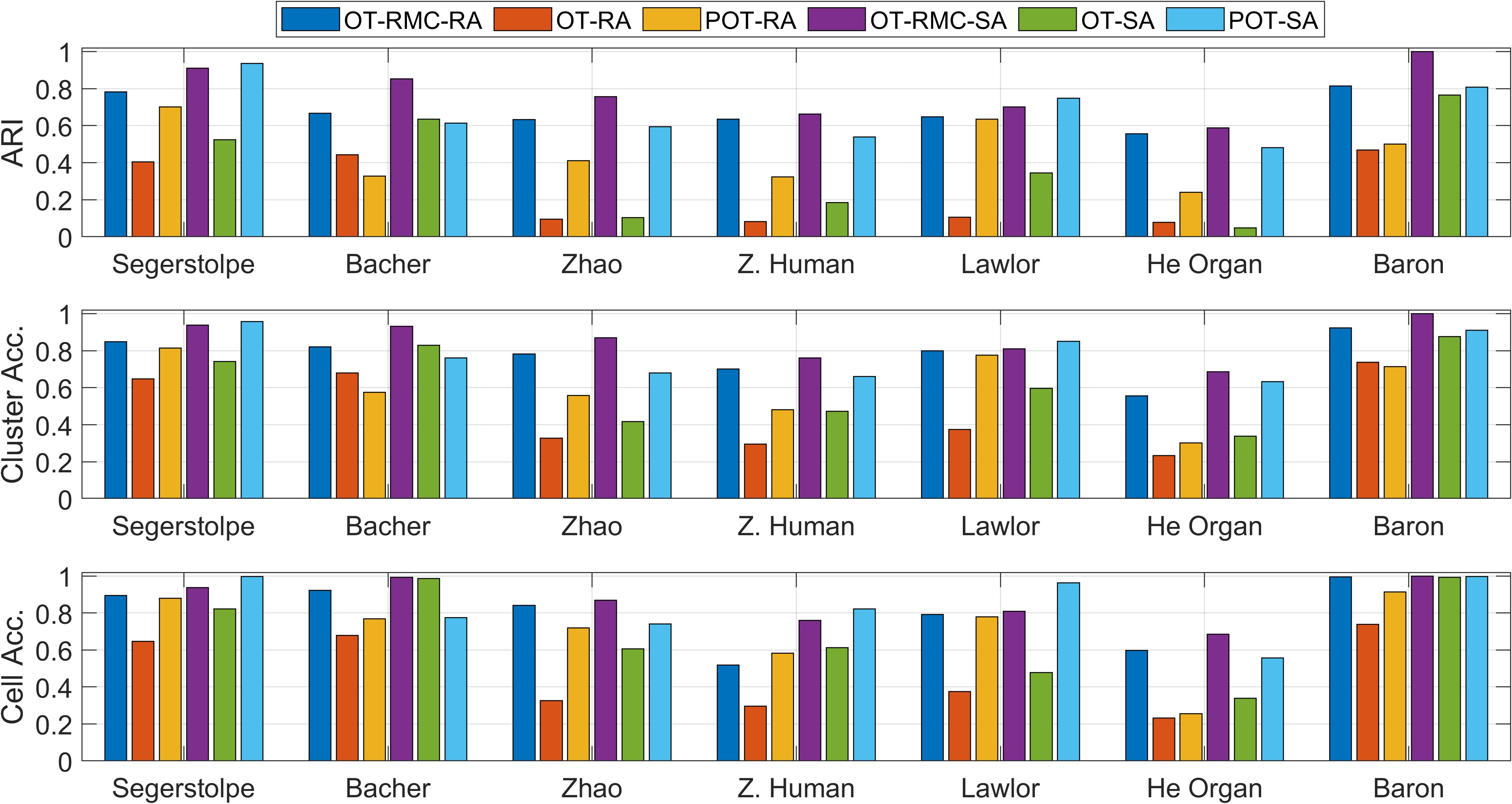}\\
        (a) Performance on datasets with real samples\\
        \vspace*{0.1in}
\includegraphics[width=\textwidth,height=0.22\textheight,keepaspectratio]{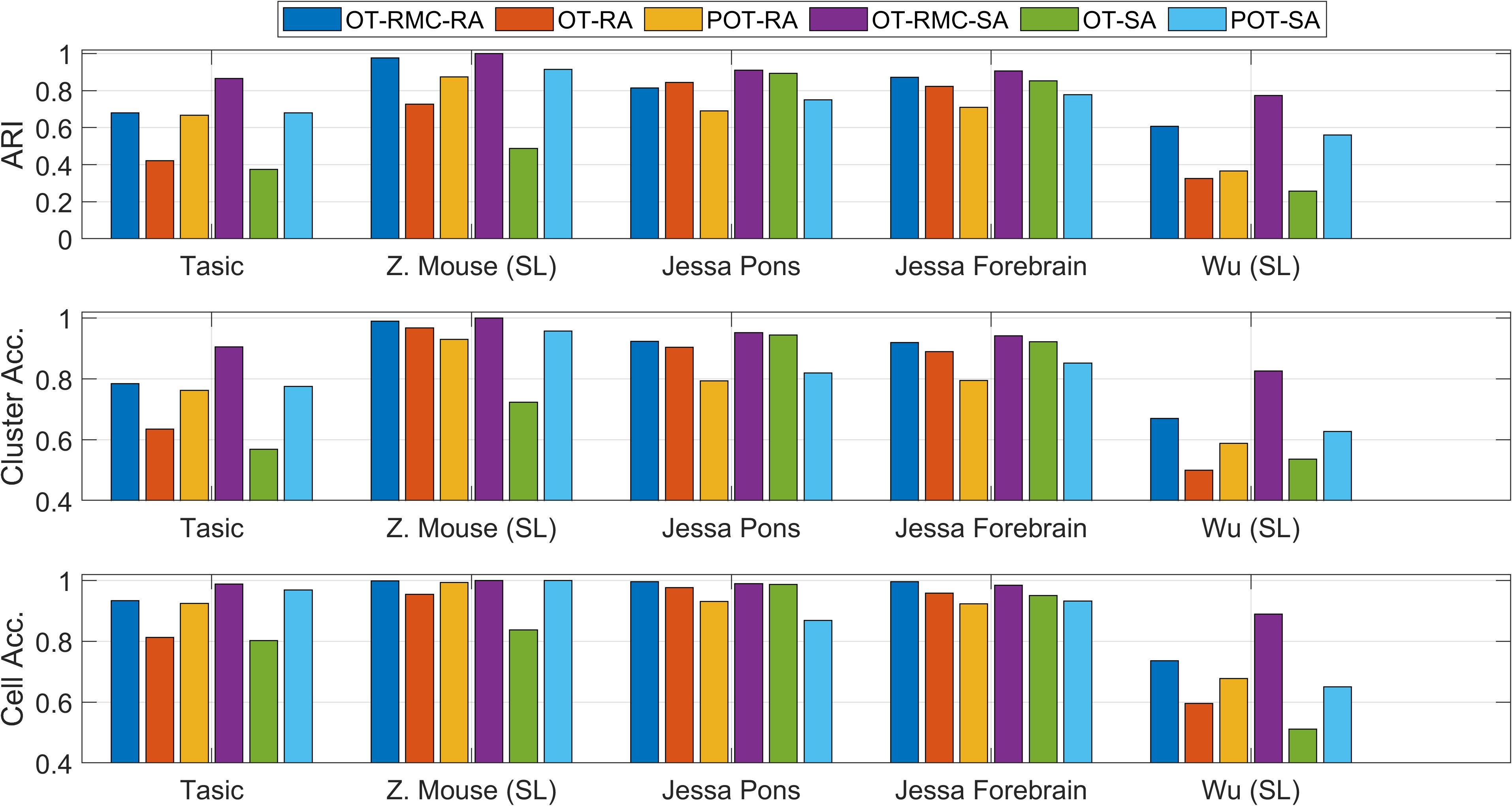}\\
        (b) Performance on datasets with simulated samples\\
         \vspace*{0.1in}
\includegraphics[width=\textwidth,height=0.22\textheight,keepaspectratio]{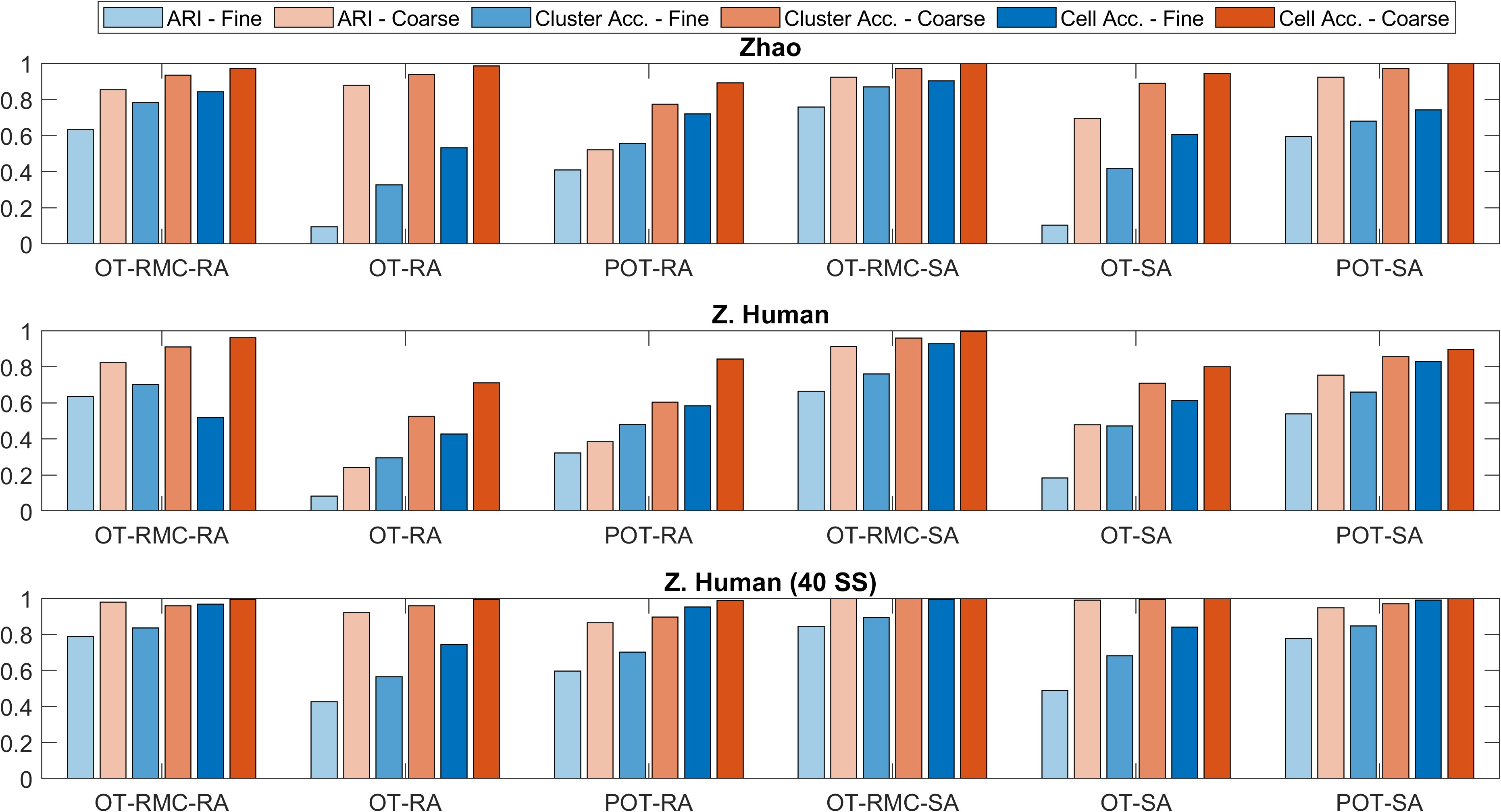}\\
      (c) Comparison of Performance: fine versus coarse cell types 
      {\spacingset{1}
    \caption{Accuracy achieved by four methods for identifying cell types based on the taxonomy. OT-RMC-SA corresponds to the MOTT system. The accuracy is measured by ARI, cluster-level accuracy $\zeta_{cls}$, and cell-level accuracy $\zeta_{cell}$.  ``SS'' stands for simulated samples. The number of simulated samples in a dataset has a default value of 20 unless specified in the parenthesis. If ``SL'' is indicated, the samples were simulated by randomly dividing cells within different cell conditions separately; otherwise, the samples were generated by randomly dividing the entire data. Z. Human stands for Zilionis Human and Z. Mouse stands for Zilionis Mouse. Some datasets had two sets of ground truth labels; coarse indicates major cell types and fine cell subtypes.}
    \label{Fig4}
    }
\end{figure}

To numerically assess the taxonomy results, we calculate the cluster-level accuracy $\zeta_{cls}$ and cell-level accuracy $\zeta_{cell}$ in terms of labeling cell clusters. Fig~\ref{Fig4} presents the results achieved by the four methods: OT-RMC-SA, OT-SA, OT-RMC-RA, and OT-RA. 
When RA is used, the results shown are average values over all valid reference samples, specifically those containing the maximum number of cell clusters.
For clarity, we present the results obtained from datasets with real samples in one plot (Fig~\ref{Fig4}(a)) and those from simulated samples in another plot (Fig~\ref{Fig4}(b)). In Fig~\ref{Fig4}(a), we use fine ground truth labels for cell types in the Zhao and Zillions Human datasets. To evaluate the impact of coarse versus fine labels on the taxonomy's performance, we compare the results obtained from these two sets of labels in Fig~\ref{Fig4}(c).

As shown in Fig~\ref{Fig4}(a) and (b), MOTT (i.e., OT-RMC-SA) consistently outperforms the other methods across both real and simulated sample datasets, regardless of the evaluation measures used. The only exceptions are for the Segerstolpe and Lawlor pancreas data where POT-SA does only slightly better than MOTT.  In most cases when MOTT performs best, OT-RMC-RA follows as the second-best performing method, followed by POT-SA. 
When comparing the cluster matching schemes, OT-RMC versus OT, the former consistently yields better results than the latter in both RA and SA cases, with the disparity in performance often being pronounced. This indicates that the marginal constraints imposed by OT can be detrimental for these datasets, which frequently include samples with different cell types, as OT fundamentally eliminates the possibility of new clusters. When comparing the cluster alignment schemes, SA versus RA, OT-RMC-SA almost always outperforms OT-RMC-RA, with the difference being remarkable for many datasets. This suggests that it is advantageous to simultaneously align clusters rather than performing alignment with a pre-selected reference. Fig~\ref{Fig4}(a) and (b) also demonstrate that building a taxonomy with OT-RMC or POT often significantly improves performance. Indeed the only instance when a taxonomy does not help is with the Jessa Pons data (having simulated samples) and using POT.

The type of dataset significantly affects the performance of the methods. For instance, the performance gain of OT-RMC-SA is weakest and marginal on the He Organ dataset. This dataset is unique in that the samples correspond to different human organs from a single patient, in contrast to samples from different patients. Additionally, the four methods perform better on simulated samples than on real samples. One likely reason is that the simulation of samples sidesteps the issue of batch effects~\cite{luo}. 
As shown by Fig~\ref{Fig4}(b), unlike the real sample datasets, and the Tasic and Wu datasets, Z. Mouse, Jessa Pons and Jessa Forebrain yield less pronounced performance differences among the methods. Except for the Jessa Pons and Jessa Forebrain datasets, OT-SA yielded the worst performance across all three metrics by significant margins.

In Fig~\ref{Fig4}(c), we investigate the impact of smaller cluster sizes on the performance. Specifically, for three datasets, there are two sets of cell type labels: coarse versus fine. The fine cell types are obtained by dividing coarse cell types into subgroups. As a result, cell clusters are smaller under the fine type labels. These three datasets are the Zhao, Zilionis Human, and Zilionis Human SS-40. The Zilionis Human SS-40 dataset contains the same cells from the Zilionis Human dataset but the division into 40 samples is randomly generated. In all three datasets we see the performance across the three metrics being higher when coarse cell types are used. Fig~\ref{Fig4}(c) also shows that the gap in performance between the fine and coarse cell types to be much greater for the OT-RA and OT-SA methods relative to the other methods. For all three datasets we observe that OT-RMC-SA and POT-SA are the least sensitive methods, with OT-RMC-SA notably more robust on the Zhao dataset.

\begin{figure}[ht]
    \centering
    \includegraphics[width=0.75\textwidth]{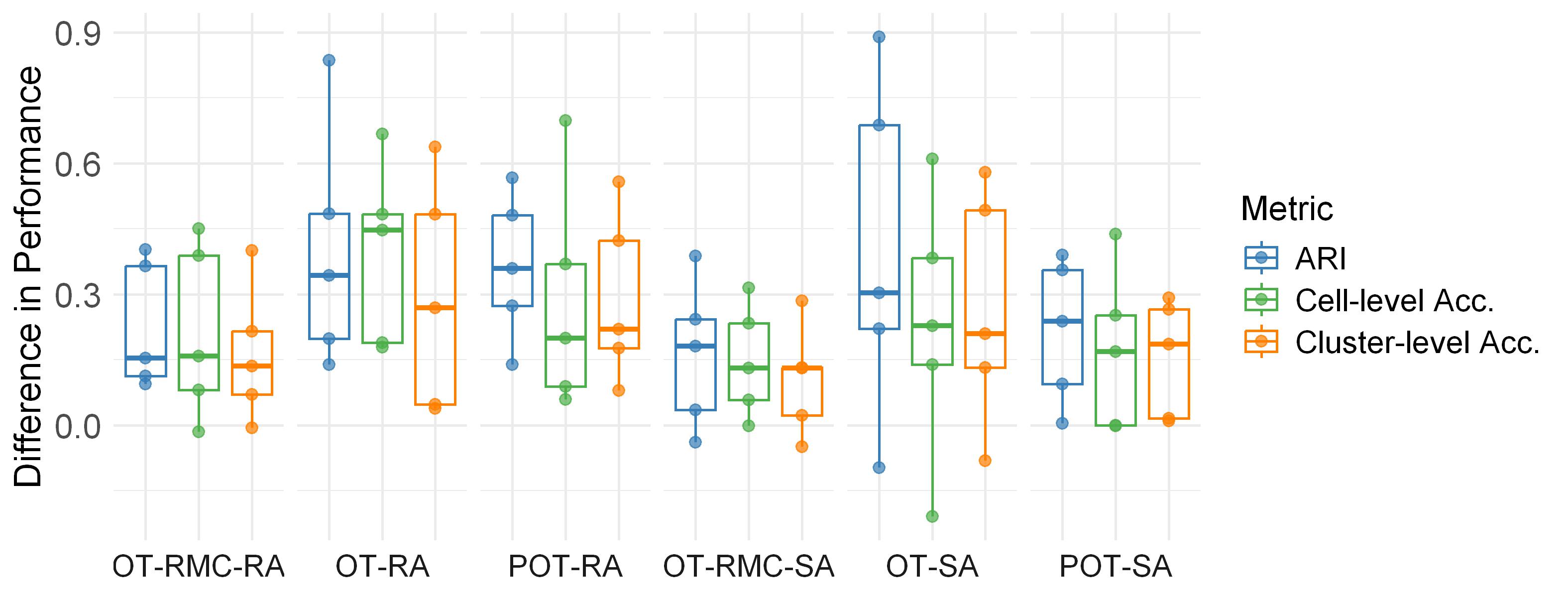}
    {\spacingset{1} 
    \caption{Boxplots showing the differences in performance between pairs of real sample and simulated sample datasets. Five datasets are analyzed: Segerstolpe, Zhao (fine cell types), Zilionis Human (fine cell types), He Organ, and Baron. Each simulated dataset (generated by the SS scheme) consists of 20 samples, except for the Zilionis Human dataset, which consists of 40 samples. Performance is evaluated using three metrics: ARI, cluster-level accuracy $\zeta_{cls}$, and cell-level accuracy $\zeta_{cell}$.
    A positive difference indicates better performance on the simulated sample datasets. In nearly all cases across different dataset pairs and methods, the simulated sample datasets show superior performance compared to their real sample counterparts. Notably, OT-RMC-RA and OT-RMC-SA exhibit smaller medians and ranges, indicating that these methods are less sensitive to how samples are generated.}
    \label{Fig5}
    }
\end{figure}

It is interesting to examine whether OT-RMC-SA is more robust than other methods when samples in a dataset deviate from ideal conditions. As previously mentioned, simulated samples created from one dataset represent ideal samples, with variation attributed only to randomness rather than any systematic cause. However, non-ideal conditions in real samples arise due to the presence of batch effects and small cell populations. We formed five pairs of datasets for comparison. Each pair consists of a real sample dataset and a simulated sample dataset created from the same source, such as the Segerstolpe dataset and its version with 20 simulated samples. We computed the differences in ARI, cell-level accuracy $\zeta_{cell}$, and cluster-level accuracy $\zeta_{cls}$ between the results from the simulated and real sample datasets. A positive difference indicates better performance on the simulated data. In Fig~\ref{Fig5}, the boxplots illustrate the differences obtained from the five pairs of datasets for each method and each performance measure. We observe that for all methods, the performance difference is predominantly positive. This aligns with our expectation: it is easier to construct a taxonomy when variation between samples is solely due to randomness. Additionally, we note that OT-RMC-SA, OT-RMC-RA and POT-SA are substantially less sensitive to the different ways samples are generated, with OT-RMC-SA performing slightly more robustly than OT-RMC-RA and POT-SA according to the median and range of $\zeta_{cell}$ and $\zeta_{cls}$.


\begin{figure*}[ht]
\centering
\begin{tabular}{ccc}
    \includegraphics[width=0.3\textwidth]{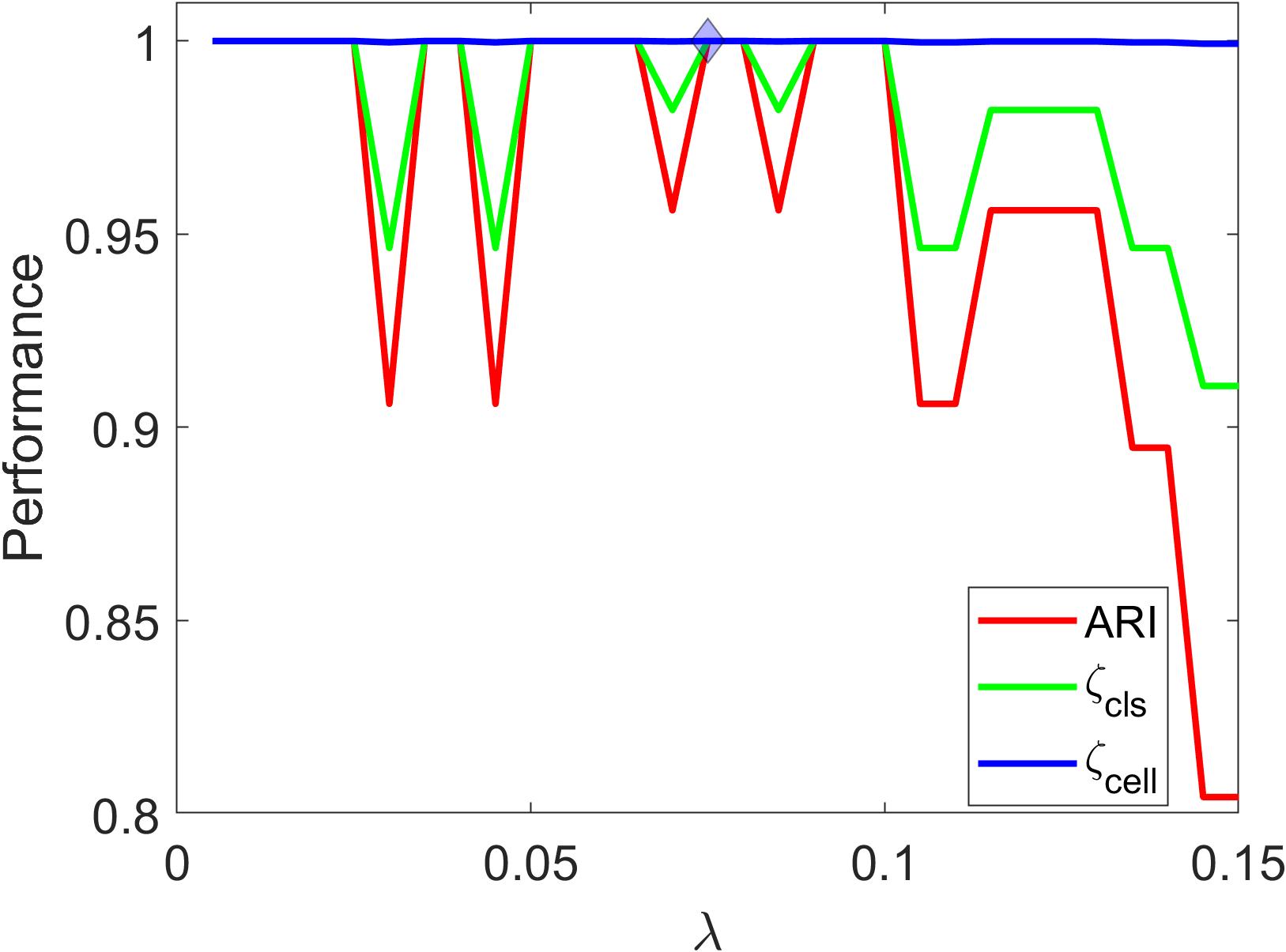} &
    \includegraphics[width=0.3\textwidth]{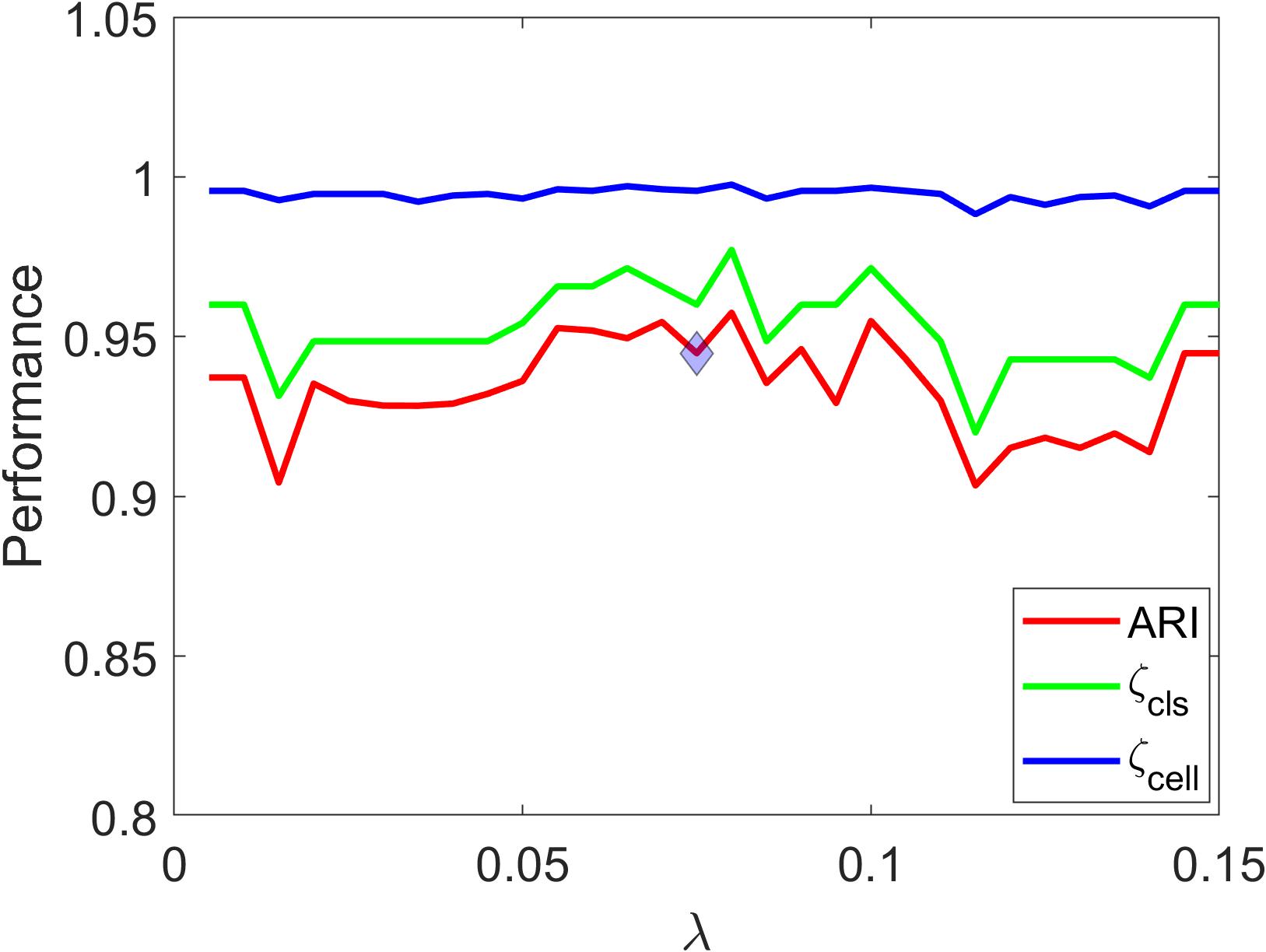} &
    \includegraphics[width=0.3\textwidth]{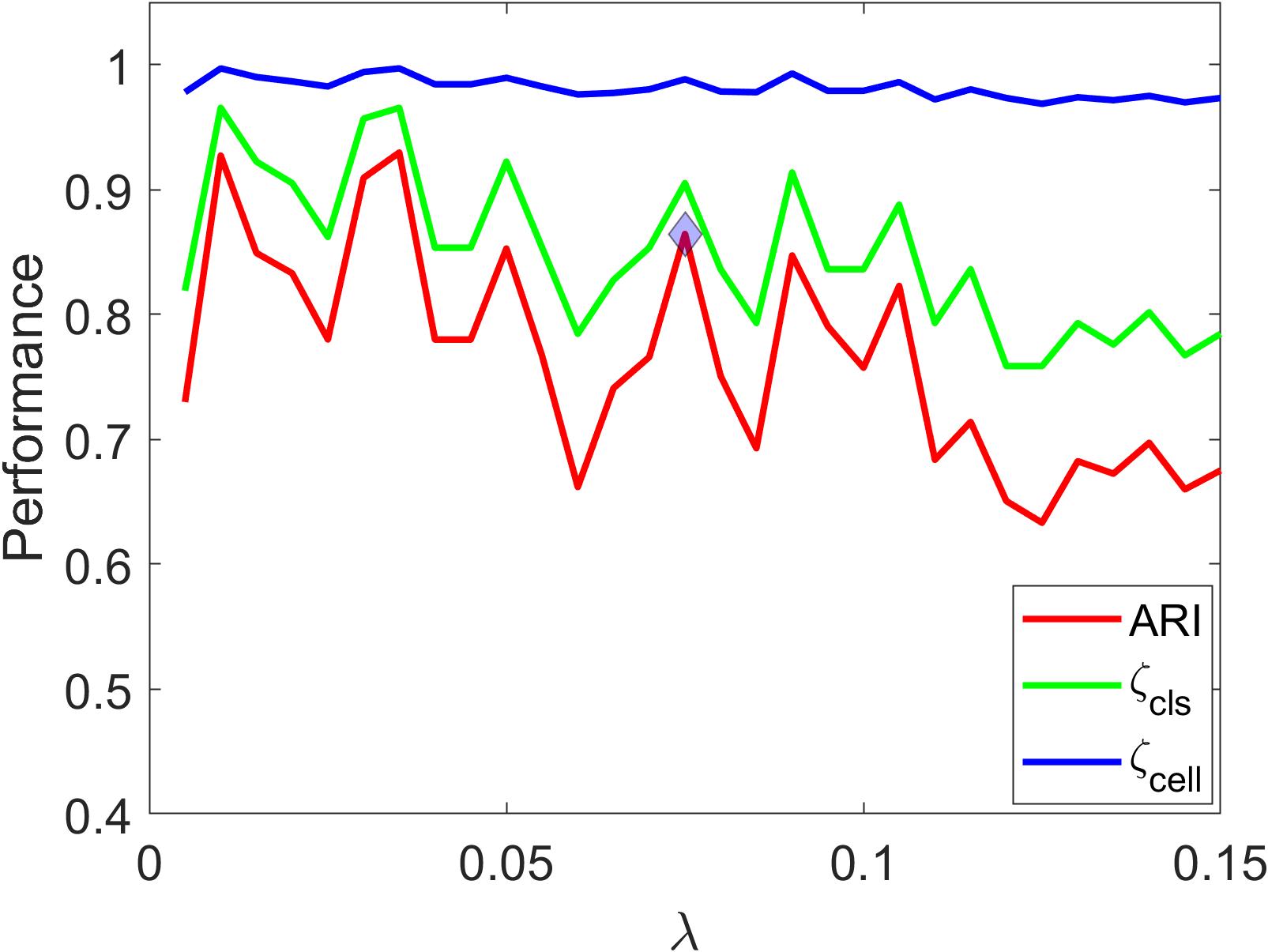} \\
        (a) Baron  & (b) Segerstolpe  & (c) Tasic 
\end{tabular}
{\spacingset{1} \caption{Cell type labeling performance achieved by OT-RMC-SA, as measured by cluster-level accuracy $\zeta_{cls}$, cell-level accuracy $\zeta_{cell}$, and ARI, with varying $\lambda$. Three datasets are analyzed: (a) Baron, (b) Segerstolpe, (c) Tasic.}
\label{Fig6}
}
\end{figure*}

\begin{table}[htbp]
\centering
\begin{tabular}{ |p{2.4cm}|P{1.4cm}|P{1.4cm}|P{1.4cm}|P{1.4cm}|P{1.4cm}|}
\hline
 Method & Metric& Segers. & Bacher & Z. Mouse &Wu\\
\hline
 \multirow{2}{*}{OT-RMC-RA}& Acc.  &.474 &.529  &.931 & .947\\
 &
 AUC &.522&.507& .939& .95\\
 \hhline{|=|=|=|=|=|=|}
 \multirow{2}{*}{OT-RA}& Acc.  &.711 &.449  &.951 & .947\\
 &
 AUC &.522&.507& .939& .95\\
 \hhline{|=|=|=|=|=|=|}
 \multirow{2}{*}{POT-RA}& Acc.  &.579 &.486  &.938 & .974\\
 &
 AUC &.583&.414& .938& 1\\
 \hhline{|=|=|=|=|=|=|}
 \multirow{2}{*}{OT-RMC-SA}& Acc.  &.9 &.6  &1 & 1\\
 &
 AUC &.99&.526& 1& 1\\
 \hhline{|=|=|=|=|=|=|}
 \multirow{2}{*}{OT-SA}& Acc.  &.85 &.5  &.95 & .95\\
 &
 AUC &.87&.507& .939& .95\\
 \hhline{|=|=|=|=|=|=|}
 \multirow{2}{*}{POT-SA}& Acc.  &.9 &.45  &1 & 1\\
 &
 AUC &.98&.604& 1& 1\\
 \hhline{|=|=|=|=|=|=|}
 \multirow{2}{*}{Ground Truth}& Acc.  &.85 &.6  &1 & 1\\
 &
 AUC &.98&.531& 1& 1\\
 \hline
\end{tabular}
{\spacingset{1} \caption{Sample-level classification accuracy on three datasets. Except for Bacher, we used simulated samples via the SL scheme. For the ``Ground Truth'' method, ground truth cell type labels are used as the taxonomy. Classification algorithm RF is applied. Accuracy is computed based on the leave-one-out scheme and is reported in terms of the percentage of correctly classified instances (Acc.) and Area Under the Curve (AUC).}
\label{classifytab}
}

\end{table}
We determine the hyperparameter $\lambda$ in OT-RMC empirically. By varying $\lambda$ from $0.005$ to $0.15$, we find that the optimal value for $\lambda$ is always around $0.075$. Hence, we set $\lambda=0.075$ for the above reported results. In Fig~\ref{Fig6}, for three datasets, we show the cell type labeling accuracy achieved by OT-RMC-SA with varying $\lambda$, as measured by $\zeta_{cls}$, $\zeta_{cell}$, and ARI. The performance is stable when $\lambda$ is not too large.

\subsection{Accuracy of Sample-level Classification}
\label{secsample}
Next, we examine the effect of taxonomy on sample-level classification using the Segerstolpe, Bacher, Zilionis Mouse, and Wu datasets. 
With the exception of the Bacher dataset, we utilized simulated samples generated through the SL scheme. As explained in the previous section, the features used for classification are the proportions of cells in each cluster. An established taxonomy ensures the coherency of cluster labels across different samples. As a performance benchmark, we also compute the proportions of cells in each ground-truth cell type and use them as classification features. In practice, ground truth labels are not expected to be available for every sample, which is the primary reason for constructing a taxonomy.

We experimented with the classification algorithms RF. Due to the small number of available samples (no greater than 20), we computed classification accuracy using the leave-one-out scheme. For scRNA-seq datasets found in the literature, although the number of cells per sample can be large, the number of samples is consistently low. The results are shown in Table~\ref{classifytab}. The results demonstrate that, compared with the other methods for building a taxonomy, OT-RMC-SA (i.e., MOTT) and POT-SA yield the most accurate classification across datasets. The advantage of OT-RMC-SA and POT-SA is most prominent with the Segerstolpe dataset. It is also notable that the ground truth taxonomy rarely outperforms OT-RMC-SA. For the Wu dataset, the classification task is straightforward, such that OT-RMC-SA, POT-SA, and the ground truth taxonomy all achieve perfect accuracy. For the Bacher data, these three methods also achieve similar classification accuracies, with POT-SA slightly worse than the other two. However, POT-SA has a higher AUC compared to the other two methods. For the Segerstolpe dataset, however, OT-RMC-SA and POT-SA perform better than the ground truth taxonomy. It is difficult to pinpoint why OT-RMC-SA or POT-SA can outperform the ground truth taxonomy. It is possible that while constructing a taxonomy, OT-RMC-SA and POT-SA capture valuable information for subsequent classification.  

\section{Discussion}
\label{con}
In this paper, we introduced a new method, called Multisample OT Taxonomy (MOTT), for constructing a hierarchical taxonomy of cell clusters from different samples. One of the major challenges in single-cell RNA sequencing is achieving consistent labeling of cell clusters across samples due to variations caused by random data noise or batch effects. Inconsistent annotations hinder our ability to accurately understand cellular diversity and to extract meaningful features for tasks such as disease classification. MOTT addresses these challenges by creating a taxonomy that identifies which cell clusters across samples represent the same intrinsic type and reveals similarities between cell types. 
Even without employing a batch effect removal tool, MOTT performs effectively and exhibits lower sensitivity to batch effects compared to other methods. Our experiments show that MOTT accurately annotates cell clusters and offers valuable insights for downstream analysis. While POT-SA achieves comparable performance to MOTT, it is often slightly outperformed. Notably, beyond the choice of the matching method, we observe that incorporating a taxonomy enhances performance on sample-level classification.

One limitation of our study is that we have only assessed the taxonomy's ability to identify cell clusters within the same cell type. While MOTT organizes these clusters hierarchically, establishing a foundation for defining similarity measures between cell types, our data lacks ground truth for evaluating any specific definition of similarity. As a result, we have not tested the taxonomy's effectiveness in quantifying cell type similarity.

Furthermore, exploring dynamic contexts beyond static sample collections presents an interesting avenue for future work. For instance, our current approach treats samples as independent, whereas they could be related over time (e.g., samples from the same patient collected at different periods) or through other factors. Incorporating such contextual information into the cluster alignment process—using methods like OT-RMC—could enhance performance by more effectively capturing the changes in cell clusters across related samples.


\section*{Data Availability}
The source data used are in public domain in other publications in the reference. The processed data generated and analyzed during this study and the code implementation are available on Github: https://github.com/sebastian5799/Multisample-OT-Taxonomy

\section*{Acknowledgments}
J. Li's research is supported by the National Science Foundation under grant CCF-2205004.


\printbibliography

\section*{Supporting information}

\paragraph*{S1 Appendix.}
\label{S1_appendix}
{\bf This file includes a description of the basic formulation of optimal transport and detailed information about the datasets and the data processing steps.
}

\end{document}